\documentclass{svjour3}                     
\smartqed  
\usepackage{graphicx}
\usepackage{amsmath}
\usepackage{amssymb}
\usepackage{color}
\usepackage{verbatim}
\usepackage{array}
\usepackage{graphicx}
\usepackage{epsfig}
\usepackage{multirow}
\usepackage{bm}
\newcommand{\ds}{\displaystyle}
\newcommand{\ddsum}[1]{{\displaystyle \sum_{ #1 }}}

\newcommand{\supercomas}[1]{``#1''}

\newcommand{\bra}[1]{\mathinner{\langle{#1}|}}
\newcommand{\ket}[1]{\mathinner{|{#1}\rangle}}
\newcommand{\braket}[2]{\langle #1|#2\rangle}

\begin{document}

\title{Polarization correlations in the two--photon decay of hydrogen--like ions}

\author{Filippo Fratini \and Andrey Surzhykov}

\institute{Filippo Fratini \at
              Physikalisches Institut der Universit\"at Heidelberg\\
              Tel.: +49-06221549355\\
              \email{fratini@physi.uni-heidelberg.de}           
           \and
           Andrey Surzhykov \at
           Physikalisches Institut der Universit\"at Heidelberg\\
           Tel.: +49-06221549355\\
           \email{surz@physi.uni-heidelberg.de}
}

\date{Received: date / Accepted: date}

\maketitle

\begin{abstract}
Polarization properties of the photons emitted in the two--photon
decay of hydrogen--like ions are studied within the framework of the
density matrix and second-order perturbation theory. In particular,
we derive the polarization correlation function that gives the
probability of the (two--photon) coincidence measurement performed
by polarization--sensitive detectors. Detailed calculations of this
function are performed for the $2s_{1/2} \to 1s_{1/2}$ transition
in neutral hydrogen as well as Xe$^{53+}$ and U$^{91+}$ ions. The
obtained results allow us to understand the influence of
relativistic and non--dipole effects on the polarization
correlations in the bound--bound two--photon transitions in heavy
ions.
\end{abstract}

%
%
\section{Introduction}
\label{intro}

Studies on the two--photon decay of atomic hydrogen and
hydrogen--like ions have by now a long tradition both in theory and
experiment \cite{SaP98,IlU06}. Originally focused on the
total transition rates, the interest in these studies has been recently
shifted towards the energy and angular distributions as well as the
\textit{polarization properties} of the emitted photons. A series of
$\gamma$--$\gamma$ coincidence experiments were performed, for
example, to measure the polarization correlations in the $2s_{1/2}
\to 1s_{1/2}$ decay of low--$Z$ ions. When compared with theoretical
predictions, the experimental results provided an accurate test of
Bell's inequality and hidden variable theories \cite{PeD85,KlD97}.
In contrast to low--$Z$ systems, much less attention has been
paid to the polarization properties of the photons
emitted in the decay of heavy ions. Only in the light of recent
improvements in x--ray detector techniques \cite{TaS06}, the
observation of $\gamma$--$\gamma$ (polarization) correlations in
high--$Z$ domain becomes more likely within the next few years.
Beside the detailed analysis of relativistic and quantum
electrodynamics (QED) effects, these future coincidence experiments
might provide an alternative and very promising route for studying
the parity--violation phenomena in heavy atomic systems or even for
testing the symmetry violations of Bose particles \cite{FrS10}.

\medskip

In this contribution, we apply the density matrix approach in order
to analyze the spin properties of the photons emitted in the
(two--photon) decay of hydrogen--like ions. We pay special attention
to the so--called polarization correlation function that reflects
the probability of the $\gamma$--$\gamma$ coincidence measurement
performed by two polarization--sensitive detectors. The basic
relations for such a ``polarization--polarization coincidence''
scenario are given in Sections~\ref{sec:labeling} and
\ref{sec:theory}. 
In those sections, we discuss how
the polarization correlation function can be expressed in terms of
second--order transition amplitude.
Even though the
evaluation of these amplitudes is a rather complicated task due to
the summation over the complete spectrum of the system, it can be
significantly simplified within the non--relativistic dipole
approximation for the electron--photon interaction. Making use of
such approximation, we are able then to derive a simple
analytical expression for the correlation function. In Section
\ref{sec:results} we employ this expression together with the
results from our exact relativistic calculations in order to
investigate polarization correlations in the $2s_{1/2} \to 1s_{1/2}$
decay of neutral hydrogen as well as Xe$^{53+}$ and U$^{91+}$ ions.
Results of our calculations demonstrate that relativistic and
non--dipole contributions may significantly modify the
polarization correlation function and that the effect becomes most
pronounced for unequal energy sharing between the two photons. A brief
summary of our results is given finally in Section
\ref{sec:summary}.

\begin{figure}
\center
\includegraphics[width=.55\textwidth]{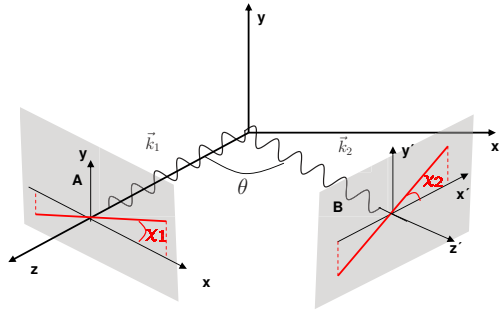}
\caption{Geometry for the two--photon decay of excited ionic state.}
\label{fig:geometry}
\end{figure}
%
%

%
%

\section{Geometry and labeling}
\label{sec:labeling}

In order to investigate the polarization correlations in the
two--photon decay of hydrogen--like ions, we shall first agree about
the geometry under which the emission of both photons is considered.
Since, for the decay of unpolarized ionic states, there is no
direction \textit{initially preferred} for the overall system, we
adopted the $z$ axis along the momentum of the ``first'' photon.
Together with the momentum direction of the ``second'' photon, this axis then
defines also the reaction plane (x–-z plane). A single polar angle
$\theta$, the so--called \textit{opening angle}, is required,
therefore, to characterize the emission of the photons with respect
to each other (cf. Fig.~1).

\medskip

As required by Bose--Einstein statistics, the two--photon state has
to be symmetric upon exchange of the particles. Therefore, it is
\textit{a priori} not possible to address them individually. We can
safely assume, however, an experimental setup in which two detectors
observe (in coincidence) the photons having certain energies and
propagation directions. A ``click'' at these detectors would
correspond to the photon's collapse into energy and momentum
eigenstates \cite{TiM09,FrT10}. A clear identity can be given,
therefore, to the photons: the {\it first (second)} photon is that
one detected by the detector $A(B)$ (marked gray in
Fig.~\ref{fig:geometry}) at a certain energy $\omega_{1(2)}$ and
momentum $\bm{k}_{1(2)}$.

\medskip

For the theoretical analysis below we shall take into account not
only the emission angles and the energies but also the
\textit{linear polarization} states of the emitted photons. In order
to observe these states, we assume that both detectors act as linear
polarizers whose transmission axes are defined in the planes that
are perpendicular to the photon momenta $\bm{k}_1$ and $\bm{k}_2$
and are characterized by the angles $\chi_1$ and $\chi_2$ with
respect to the reaction (x--z) plane. In this notation, $\chi_{1(2)} =
0^\circ$ denotes the polarization direction of the {\it first (second)} photon within the reaction
plane.

%
%
\section{Theory}
\label{sec:theory}

Within the framework of the second--order perturbation theory, the
analysis of the total as well as differential two--photon decay
rates is usually traced back to the evaluation of the transition
amplitude \cite{DrG81,GoD81}:
\begin{equation}
\begin{array}{l l l}
   \ds\mathcal{M}_{fi}^{\bf k_1\bf k_2}(\lambda_1,\lambda_2)&=&  \ds\ddsum{\nu}\!\!\!\!\!\!\!\!\int
   \left[
   \frac{ \bra{f} {\vec\alpha}\cdot\bf u_{\lambda_1}^*e^{-i\bf k_1\cdot \bf r}
   \ket{\nu}\bra{\nu} {\vec\alpha}\cdot\bf u_{\lambda_2}^*e^{-i\bf k_2\cdot \bf r\,'}\ket{i}}{E_{\nu}
   -E_i+\omega_2 }  \right.\\
   && \left. \ds +\frac{ \bra{f} {\vec\alpha}\cdot\bf u_{\lambda_2}^*e^{-i\bf k_2\cdot \bf r}
   \ket{\nu}\bra{\nu} {\vec\alpha}\cdot\bf u_{\lambda_1}^*e^{-i\bf k_1\cdot \bf r\,'}\ket{i}}{E_{\nu}
   -E_i+\omega_1 } \right] \, ,
\end{array}
\label{eq:Mfi}
\end{equation}
where $\braket{\bm{r}}{i} = \psi_{n_i j_i \mu_i}(\bm{r})$ and
$\braket{\bm{r}}{f} = \psi_{n_f j_f \mu_f}(\bm{r})$
are the well--known solutions of the Dirac-Coulomb equation for a single electron in the
initial and final states, correspondingly. In this expression, moreover, the transition
operator $\vec\alpha \, \vec u_{\lambda_j} \, e^{i \vec k_j\cdot\vec r}$ describes the relativistic
electron--photon interaction with $\vec u_{\lambda_j}$ and $\vec k_{j}$ being
the polarization and the wave vector of the $j$--th photon.

\medskip

The evaluation of the amplitude (\ref{eq:Mfi}) is not a simple task
owing to the intermediate--state summation that includes not only the
summation over the discrete part of the spectrum but also an
integration over the positive as well as the negative--energy
continuum. A number of methods have been proposed in the past to
carry out such a summation \cite{DrG81}. In the present work,
the second--order transition amplitude (\ref{eq:Mfi}) is
evaluated by means of the Green's function approach. Since this
approach has been widely applied over the last years for the
analysis of the total two--photon decay rates as well as the
polarization and angular correlation functions \cite{SuR09,RaS08},
it will not be recalled here.

\medskip

After a brief discussion of the second--order transition amplitude
(\ref{eq:Mfi}), we are now ready to analyze the polarization
properties of the emitted photons. Most naturally, such an analysis
can be performed in the framework of the density matrix theory. For
the decay of an unpolarized initial state $\ket{n_i j_i}$ into the
level $\ket{n_f j_f}$, the two--photon spin density matrix reads as:
\begin{eqnarray}
   \ds\bra{\vec k_1,\lambda_1,\vec k_2,\lambda_2}\hat\rho_f\ket{\vec k_1,\lambda_1',\vec k_2,\lambda_2'}
   &=& \frac{1}{2j_i+1} \, \sum_{\mu_i,\mu_f} \mathcal{M}_{fi}^{\vec k_1\vec k_2}(\lambda_1,\lambda_2)
   \mathcal{M}_{fi}^{\vec k_1\vec k_2\,*} (\lambda_1',\lambda_2') \, ,
\label{eq:dm}
\end{eqnarray}
where $\lambda_j = \pm1$ are the spin projections of the photons
onto their propagation directions (i.e. the so--called helicity).
Instead of this \textit{helicity} representation, it might be more
convenient to re--write the density matrix (\ref{eq:dm}) in the
representation of the vectors $\vec u_{x}$ and $\vec u_{y}$. Such vectors
denote the linear polarization of the photons respectively under the angles $\chi=0^\circ$ and
$\chi=90^\circ$ with respect to the reaction plane (see Fig.~\ref{fig:geometry}).
Any linear polarization which is nowadays measured in experiments can be expressed in terms of
these two (basis) vectors
\begin{equation}
   \vec u_{\chi} = 
   \frac{1}{\sqrt{2}}\, \left( \cos\chi \vec u_{x} +
   \sin\chi \vec u_{y} \right) = \frac{1}{\sqrt{2}} \, \left( {\rm e}^{-i\chi} \vec u_{\lambda=+1} +
   {\rm e}^{+i\chi} \vec u_{\lambda=-1} \right) \, ,
   \label{eq:cirtilin}
\end{equation}
by following the standard decomposition of linear polarization vectors in terms of the circular polarization ones.

\medskip

The density matrix (\ref{eq:dm}) still contains complete information
on two photons and, hence, can be employed to derive their
polarization properties. To achieve this, it is convenient to define
the so--called ``detector operator'' that projects out all those
quantum states of the final--state system which lead to a ``count''
at the detectors. Since in our present work we wish to analyze the
correlated linear polarization states of the photons, the detector
operator can be written as:
\begin{eqnarray}
   \ds\hat P &=& \ket{\vec k_1\chi_1 \, \vec k_2\chi_2}
   \bra{\vec k_1\chi_1 \, \vec k_2\chi_2}  \, .
\label{eq:detop}
\end{eqnarray}
From this projector operator, by taking the trace over its product
with the density matrix (\ref{eq:dm}) and applying
Eq.~(\ref{eq:cirtilin}), we immediately derive the
polarization--polarization correlation function:
\begin{eqnarray}
   \ds \Phi_{\chi_1,\chi_2}(\theta) &=& \mathcal{N}\,{\rm Tr}(\hat{P} \hat{\rho}_f) =
   \frac{\mathcal{N}}{4 (2j_i +1)} \,
   \sum\limits\limits_{\lambda_1 \lambda'_1 \lambda_2 \lambda'_2}
   \nonumber \\
   &\times&
   {\rm e}^{i (\lambda_1 - \lambda'_1) \chi_1} \,
   {\rm e}^{i (\lambda_2 - \lambda'_2) \chi_2} \,
   \mathcal{M}_{fi}^{\vec k_1\vec k_2}(\lambda_1,\lambda_2)
   \mathcal{M}_{fi}^{\vec k_1\vec k_2\,*} (\lambda_1',\lambda_2') \, ,
   \label{eq:corrfunction}
\end{eqnarray}
which represents the normalized
probability of coincidence measurement of
two photons with well--defined wave vectors $\vec k_{1}$ and $\vec
k_{2}$ and with linearly polarization vectors characterized by the
angles $\chi_1$ and $\chi_2$ with respect to reaction plane.
Here the normalization constant
$\mathcal{N}$ is chosen in such a way that, for any 
value of the opening angle
$\theta$, we get the unity
after having summed over the probabilities of the (four) independent photons' 
polarization states $\ket{xx}$, $\ket{xy}$, $\ket{yx}$, $\ket{yy}$.  
For the sake of brevity, we have introduced the notation $\ket{xx} =
\ket{\chi_1 = 0^\circ}  \ket{\chi_2 = 0^\circ}$,
$\ket{xy} = \ket{\chi_1 = 0^\circ} \ket{\chi_2 =
90^\circ}$ and so forth.

\medskip

Any further evaluation of the polarization correlation
(\ref{eq:corrfunction}) requires, in general, a computation of the
fully--relativistic transition amplitude (\ref{eq:Mfi}). This
amplitude accounts for the \textit{full} interaction $\vec\alpha \,
\vec u_{\lambda} \, e^{i \vec k\cdot\vec r}$ between the electron
and the radiation field and, hence, includes the higher non--dipole
effects. The non--dipole corrections, however, are usually expected
to be negligible for low--$Z$ ions. For these ions, it is therefore
justified to threat the electron--photon interaction within the
non--relativistic dipole approximation by setting $e^{i \vec k
\cdot\vec r} \approx 1$. Within such dipole approximation, a simple
\textit{analytic} expression for the function $\Phi$ can be
obtained:
\begin{equation}
   \ds \Phi^{2E1}_{\chi_1,\chi_2}(\theta) = \frac{1}{1+\cos^2\theta}\Big(
   \sin\chi_1\sin\chi_2 +
   \cos\chi_1\cos\chi_2\cos\theta \Big)^2 \, ,
   \label{eq:pcfget}
\end{equation}
for the particular case of the $2s_{1/2} \to 1s_{1/2}$ two--photon transition.

\begin{figure*}
\begin{center}
\vspace{.2cm}
\includegraphics[width=0.8\textwidth]{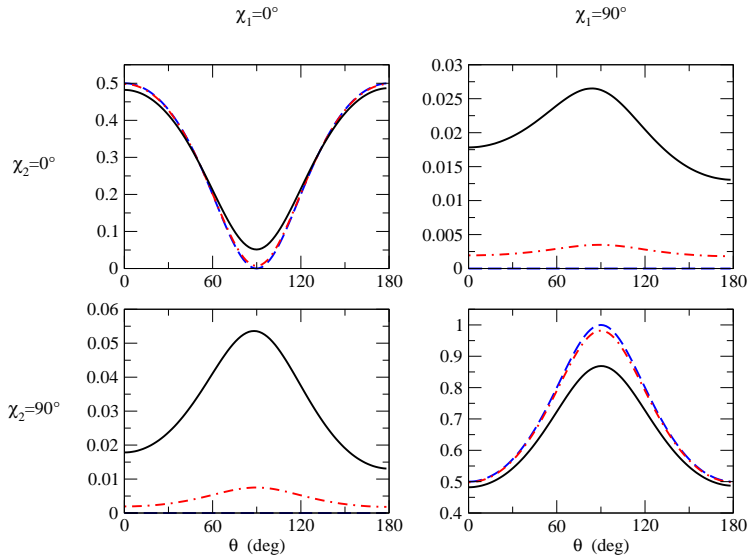}
\caption{(Color online) Polarization--polarization correlations 
of photons emitted
in
the $2s_{1/2} \to 1s_{1/2}$ two--photon decay. 
The solid line corresponds to U$^{91+}$, the dot--dashed line corresponds to Xe$^{53+}$ while
the dashed line corresponds {\it both} to H {\it and} to the non--relativistic calculation, 
since the discrepancy between them less is than $10^{-10}$\%.
Calculations are presented for the energy sharing $x$ =
1/16 and for the different angles $\chi_i$ of the (polarization)
transmission axes of detectors (cfr. Fig.~\ref{fig:geometry}).
} 
\label{fig:x_16}
\end{center}
\end{figure*}
%
%

%
%
\section{Results and discussion}
\label{sec:results}

In the present work, we have employed both the non--relativistic
electric--dipole approximation (\ref{eq:pcfget}) and the exact
relativistic treatment from Eq.~(\ref{eq:corrfunction}) in order to
investigate the polarization correlations between the photons
emitted in the $2s_{1/2} \to 1s_{1/2}$ decay. For this bound--bound
transition, calculations have been performed for neutral hydrogen atom H,
hydrogen--like xenon Xe$^{53+}$ and uranium U$^{91+}$ ions,
for various values of the \textit{energy sharing} parameter.
This parameter reflects the fraction of energy carried away by the
\supercomas{first} photon: $x = \omega_1/(\omega_1 + \omega_2)$ and,
hence, is defined in the range $0 \le x \le 1$. In Fig.~2, for
example, the polarization correlation function $\Phi(\theta)$ is
displayed for the parameter $x$ = 1/16. As seen from the figure, for
light ions,
both the electric dipole and the
fully relativistic treatments basically coincide and are well
described by Eq.~(\ref{eq:pcfget}). In particular, the polarization
correlation function $\Phi$ almost vanishes if the
polarization axes of detectors are perpendicular to each other:
$\chi_1$ = 0$^\circ$ and $\chi_2$ = 90$^\circ$ \textit{or} $\chi_1$
= 90$^\circ$ and $\chi_2$ = 0$^\circ$. Such a behaviour can be
understood if we recall that ---within the non--relativistic electric
dipole approximation--- photons are emitted in a \textit{pure} spin
state described by the state vector \cite{FrT10}:
\begin{equation}
   \label{eq:pure_state}
   \ket{\Psi_{2\gamma}}^{2E1} = \frac{1}{1 + \cos^2\theta} \,
   \left( \ket{yy} + \cos\theta \ket{xx} \right) \, .
\end{equation}
From this expression, it immediately follows  that
the probability of measuring --in coincidence-- 
orthogonal linear polarizations of photons is zero.

\medskip

For high--$Z$ hydrogen--like ions,
we expect that Eqs.~(\ref{eq:pcfget})--(\ref{eq:pure_state}) might not describe well the polarization properties of 
the emitted photons, owing to
relativistic and non--dipole effects. As seen from Fig.~2, these
effects result in a non--vanishing correlation function $\Phi$ for
the ``perpendicular polarization'' measurements, i.e. when $\chi_2=\chi_1\pm90^\circ$. 
The probability of ``parallel polarization'' measurements consequently 
decreases of the same measure.
We moreover notice that, in case of energy sharing $x$ = 1/16 and $\theta = 90^\circ$,
events with photon polarizations \textit{within} the reaction plane
are not forbidden, in contrast to the non--relativistic dipole approximation, as a consequence of relativistic and non--dipole effects. 
This, in turn, leads to a further reduction of the probability for those
events with photon polarizations which are perpendicular to the reaction plane.
For hydrogen--like uranium ions
U$^{91+}$, for example, the function $\Phi_{\chi_1,\chi_2}$, as calculated at
$\chi_1 = \chi_2 = 90^\circ$ and perpendicular photon emission
($\theta = 90^\circ$), decreases from 1 to almost 0.85 if the higher
multipole terms are taken into account.

\medskip

While the relativistic and retardation effects are significant in
high--$Z$ domain if one of the photons is much more energetic than
the second one, they become almost negligible if the photons are
emitted with nearly the same energy: $\omega_1 \approx \omega_2$.
As can be seen from Fig.~3,
for energy sharing $x$ = 0.5, the polarization probabilities obtained within
non--relativistic electric--dipole and rigorous
relativistic approaches
differ in fact only of about $\sim10^{-3}$, even for the decay of hydrogen-like uranium ions.
\begin{figure*}
\begin{center}
\vspace{.2cm}
\includegraphics[width=0.8\textwidth]{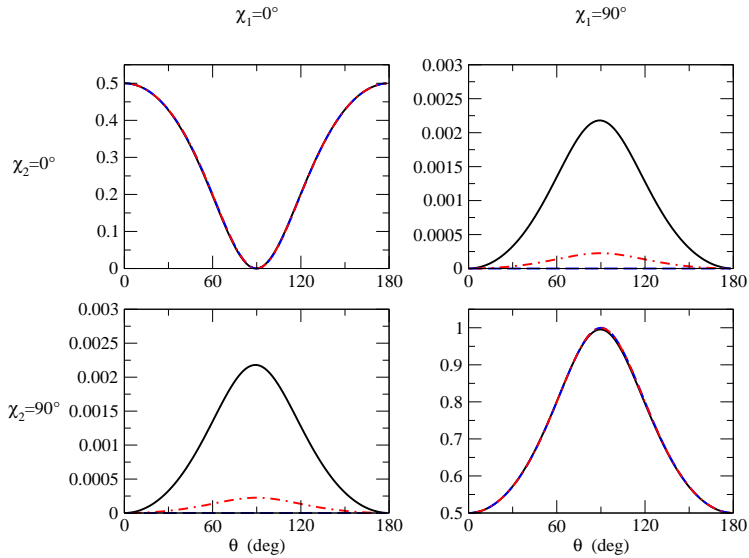}
\caption{(Color online) Polarization--polarization correlations 
of photons emitted
in
the $2s_{1/2} \to 1s_{1/2}$ two--photon decay. 
The solid line corresponds to U$^{91+}$, the dot--dashed line corresponds to Xe$^{53+}$ while
the dashed line corresponds {\it both} to H {\it and} to the non--relativistic calculation, 
since the discrepancy between them less is than $10^{-10}$\%.
Calculations are presented for the energy sharing $x$ =
1/2 and for the different angles $\chi_i$ of the (polarization)
transmission axes of detectors (cfr. Fig.~\ref{fig:geometry}).
}
\label{fig:x_12}
\end{center}
\end{figure*}
%
%

%
%
\section{Summary and outlook}
\label{sec:summary}

In summary, the polarization of the radiation emitted in the
two-photon decay of hydrogen--like ions has been investigated within
the framework of the second--order perturbation theory. Special
attention has been paid to the correlated spin states of two photons as
can be measured in coincidence $\gamma-\gamma$ experiments. In order
to predict the outcome of such experiments, the expression for the
polarization correlation function has been derived within both the
exact relativistic theory and the non--relativistic
electric--dipole approximation. Making use of these two approaches,
the photon--photon polarization correlations have been calculated
for the $2s_{1/2} \to 1s_{1/2}$ transition in neutral hydrogen as
well as hydrogen--like Xe$^{53+}$ and U$^{91+}$ ions. 
As seen from
the results obtained, the higher non--dipole terms in the
electron--photon interaction may affect the correlation function by
about 10--20 \%; effect that becomes most pronounced for the decay
of high--$Z$ ions if the major fraction of the (two–-photon)
transition energy is carried away by a single photon.

\medskip

In the present work, we have restricted our theoretical analysis of
the two--photon decay to the one--electron atomic systems. For
high--$Z$ domain, however, the \textit{helium--like} ions are the
most suitable candidates for two--photon studies. In the future,
therefore, we plan to extend our theoretical approach for studying
transitions in these two--electron species. A first case study on
the polarization correlations and spin entanglement in the $1s_{1/2}
2s_{1/2}: J=0 \to 1s_{1/2}^2: J=0$ decay of U$^{90+}$ ions is
currently under way and will be published soon.

\begin{acknowledgements}
We acknowledge the support from the Helmholtz Gemeinschaft and GSI
under the project VH--NG--421.
\end{acknowledgements}


\end{document}